\documentclass{article}
\usepackage{dcolumn}
\usepackage{booktabs}
\usepackage{multirow}

\usepackage{xcolor, soul}
\usepackage[draft]{hyperref}

\usepackage{authblk}
\usepackage{setspace}
\usepackage[margin=1.25in]{geometry}
\usepackage{graphicx}
\graphicspath{ {./figures/} }
\usepackage{graphicx}
\usepackage{subfigure}
\usepackage{graphicx, epsfig, epstopdf}
\usepackage{amsmath}

\usepackage{amsmath}
\usepackage{color}
\usepackage{mathtools}

\usepackage[style=nejm, 
citestyle=numeric-comp,
sorting=none]{biblatex}
\begin{document}
\title{Transfer Learning for Inverse Design of Tunable Graphene-Based Metasurfaces}

\author[1*]{Mehdi Kiani}
\author[2]{Mahsa Zolfaghari}
\author[3]{Jalal Kiani}

\affil[1*]{Department of Electrical Engineering, Iran University of Science and Technology, Tehran, Iran.}
\affil[2]{School of Digital Technologies and Arts, Staffordshire University London, UK}
\affil[3]{Department of Civil Engineering, University of Memphis, Memphis, TN 38152, USA.}
\affil[*]{Address correspondence to: mhdkianii@gmail.com}

\date{}

\onehalfspacing

\maketitle

\begin{abstract}
This paper outlines a new approach to designing tunable electromagnetic (EM) graphene-based metasurfaces using convolutional neural networks (CNNs). EM metasurfaces have previously been used to manipulate EM waves by adjusting the local phase of subwavelength elements within the wavelength scale, resulting in a variety of intriguing devices. However, the majority of these devices have only been capable of performing a single function, making it difficult to achieve multiple functionalities in a single design. Graphene, as an active material, offers unique properties, such as tunability, making it an excellent candidate for achieving tunable metasurfaces. The proposed procedure involves using two CNNs to design the passive structure of the graphene metasurfaces and predict the chemical potentials required for tunable responses. The CNNs are trained using transfer learning, which significantly reduced the time required to collect the training dataset. The proposed inverse design methodology demonstrates excellent performance in designing reconfigurable EM metasurfaces, which can be tuned to produce multiple functions, making it highly valuable for various applications. The results indicate that the proposed approach is efficient and accurate and provides a promising method for designing reconfigurable intelligent surfaces for future wireless communication systems.
\end{abstract}


\section{Introduction}
    \begin{figure*}[t]
        \centering
        \includegraphics[height=2.3 in]{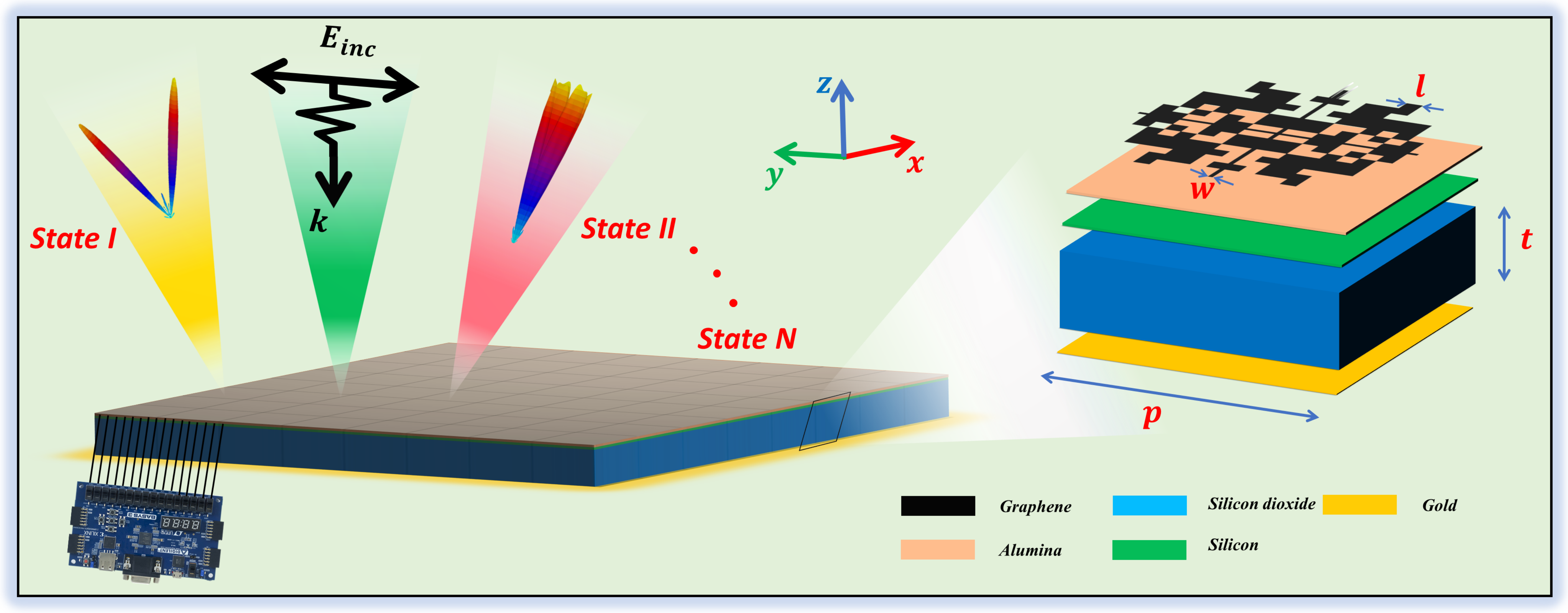}
        \caption{\label{fig:epsart} {A conceptual illustration of the reconfigurable metasurface, as well as the five-layer graphene-based meta-atoms used for training the proposed inverse design model.}}
    \end{figure*}
    Electromagnetic (EM) metasurfaces have gained significant attention due to their ability to control EM wave propagation in a subwavelength thickness \cite{chen2016review, li2018metasurfaces}. While this technology has led to various advancements, including mantle cloaking \cite{chen2011mantle, sounas2015unidirectional}, polarization twisting \cite{zhao2011manipulating, grady2013terahertz}, wave-front manipulation \cite{monticone2013full, yu2011light}, and perfect absorption \cite{landy2008perfect, tao2008metamaterial}, there are certain limitations that need to be addressed. One such limitation pertains to the inflexibility of EM characteristics, particularly in operating systems  \cite{chen2006active, chen2017reconfigurable}. Consequently, there is a growing need for reconfigurable metasurfaces that can dynamically manipulate EM waves in response to external signals.

    Reconfigurability has been achieved at microwave frequencies by incorporating semiconductor lumped components into metasurfaces \cite{li2017nonlinear, kiani2020self, kiani2020spatial}. At Terahertz (THz) frequencies, tuning substances such as vanadium dioxide \cite{hashemi2016electronically}, liquid crystal \cite{decker2013electro}, and graphene \cite{fallahi2012design} have shown promise as platforms for reconfigurable metasurfaces. Graphene, a single two-dimensional (2D) plane of carbon atoms arranged in a hexagonal lattice, exhibits substantial and configurable absorption in the THz regime \cite{ju2011graphene, grigorenko2012graphene}. The ability to control graphene's absorption behavior by changing its chemical potential through electrostatic or chemical doping makes it a promising candidate for reconfigurable THz devices \cite{chen2011controlling}.

    One critical issue in designing reconfigurable metasurfaces, including graphene-based ones, is the inverse design of tunable meta-atoms that can exhibit different desired EM responses by altering external control signals. Traditional inverse design methods involve time- and computation-intensive searches over known structures, such as cross-shaped patches, rectangular patches, and split-ring resonators, which often fall short of achieving the required performance, especially when broadband, polarization-sensitive, and wide-angle responses are desired. An evolutionary optimization algorithm was proposed as an alternative to finding optimal THz absorbers with tunable performance \cite{torabi2017evolutionary}. However, this stochastic algorithm heavily relies on the quality of initial designs, limiting its consistency and productivity as the problem complexity increases.
	
    Deep Learning (DL) approaches, on the other hand, have emerged as powerful representation-learning methods. These approaches assemble simple yet nonlinear modules that map high-dimensional structured data into lower-dimensional representations, and the layers are learned from data rather than being developed by human engineers \cite{lecun2015deep}. DL has gained popularity in various fields, including computer vision\cite{krizhevsky2012imagenet, voulodimos2018deep}, natural language processing \cite{nadkarni2011natural}, reinforcement learning \cite{krizhevsky2012advances}, graph representation learning \cite{kipf2016semi, hashemi2023visiting}
    drug discovery \cite{vamathevan2019applications}, medical diagnosis \cite{kononenko2001machine}, and different fields of engineering \cite{campbell2020explosion, kiani2020application, raccuglia2016machine, kiani2019application}.
    
    In the field of EM design, DL  approaches have demonstrated promise in designing a wide range of well-functioning EM devices by directly identifying key geometric parameters based on desired EM responses. Peurifoy et al. utilized fully connected Neural Networks (NNs) to simulate light interaction with nanoscale structures \cite{peurifoy2018nanophotonic}. Similarly, Nadell et al. employed NNs with fully connected layers along with a fast-forward dictionary search method for the inverse design of all-dielectric metasurfaces \cite{nadell2019deep}. However, the use of fully connected NNs limited the efficiency of the inverse design models to simple structure designs with restricted EM responses.
    In contrast, Convolutional Neural Networks (CNNs) have been successfully applied to design various types of metasurfaces, leading to the achievement of new functionalities and improved device performance \cite{liu2018generative, wen2020robust, so2019designing, yeung2021multiplexed}. For instance, Liu et al. developed a Generative Adversarial Network (GAN) based on CNNs for the inverse design of metasurfaces \cite{liu2018generative}. However, these metasurface design schemes typically focused on either the amplitude or phase response of the metasurfaces \cite{qiu2019deep, zhang2019machine}. To address this limitation, Naseri et al. proposed a generative DL model based on a variational autoencoder for the inverse design of multi-layer metasurfaces, considering both the amplitude and phase responses, albeit for a single function, polarization conversion \cite{naseri2021generative}.
    To tackle the challenge of generating metasurfaces with multiple functionalities, two inverse design models have been introduced. Kiani et al. utilized conditional GANs in the microwave regime to design multi-layer metal-dielectric metasurfaces with three different functions and full-space coverage \cite{kiani2022conditional}. In a similar vein, An et al. employed a combination of conditional and Wasserstein GANs for the inverse design of all-dielectric metasurfaces in photonics \cite{an2021multifunctional}.
	
	Notwithstanding, despite the advancements in DL models for metasurface design, significant restrictions still persist. Firstly, these models are limited to designing passive metasurfaces with fixed functions, which further constrains their applicability to a limited frequency range. Secondly, the data collection process associated with these models is time-consuming, resulting in overall inefficiency. As a consequence, the DL-enabled design of reconfigurable metasurfaces, especially when the training dataset is significantly reduced, presents a formidable challenge that has remained unaddressed until now.
    In this study, a novel approach based on Transfer Learning (TL) is proposed for the inverse design of reconfigurable graphene metasurfaces. The approach utilizes two consecutive CNNs that leverage TL to train the networks and facilitate the design of tunable meta-atoms. In the proposed model, the first network designs the passive components of the tunable graphene meta-atoms, while the second network predicts the chemical potentials (control signals) of the graphene meta-atoms to achieve tunable responses. To train, validate, and test the networks, datasets are constructed comprising graphene meta-atoms represented as 16$\times$16 matrices. These datasets incorporate both graphene and vacuum square blocks with chemical potentials ranging from 100 $meV$ to 1000 $meV$, along with their corresponding phase responses encompassing a wide range of values. It is demonstrated that the incorporation of pre-trained CNNs in the inverse design model improves the feature extraction of EM phase responses, yielding satisfactory results even with a limited amount of training data samples. Finally, the effectiveness of the approach is showcased by successfully employing it in the inverse design of a tunable meta-atom. This meta-atom exhibits a wide range of desired phase responses by simply adjusting the chemical potentials. The presented approach offers several unique features in comparison to previous works in the literature. Firstly, the framework enables the design of tunable graphene metasurfaces without relying on initial designs, distinguishing it from the optimization-based approach presented in \cite{torabi2017evolutionary}. Secondly, this approach represents the first endeavor to design reconfigurable graphene metasurfaces using DL-based metasurface inverse design methods. Lastly, the proposed methodology leverages the power of TL, resulting in significant reductions in computational time and the cost associated with training data collection.


%
\begin{figure}[t]
    \centering
    \includegraphics[height=2.4in]{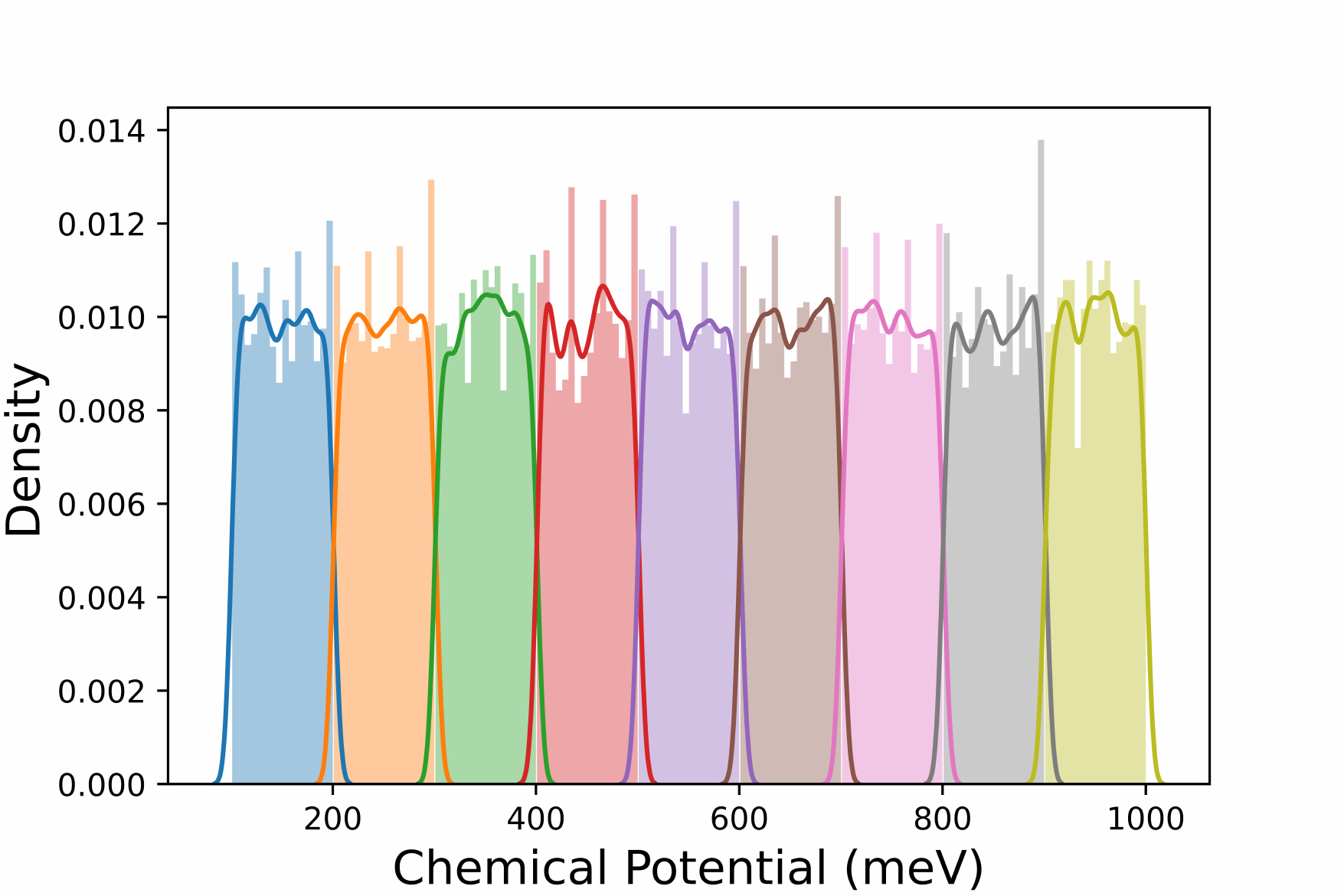}
    \caption{\label{} {The chemical potential distribution of the graphene meta-atoms in the training, validation, and test datasets.}}
\end{figure}
\begin{figure*}[t]
    \centering
    \includegraphics[height=4in]{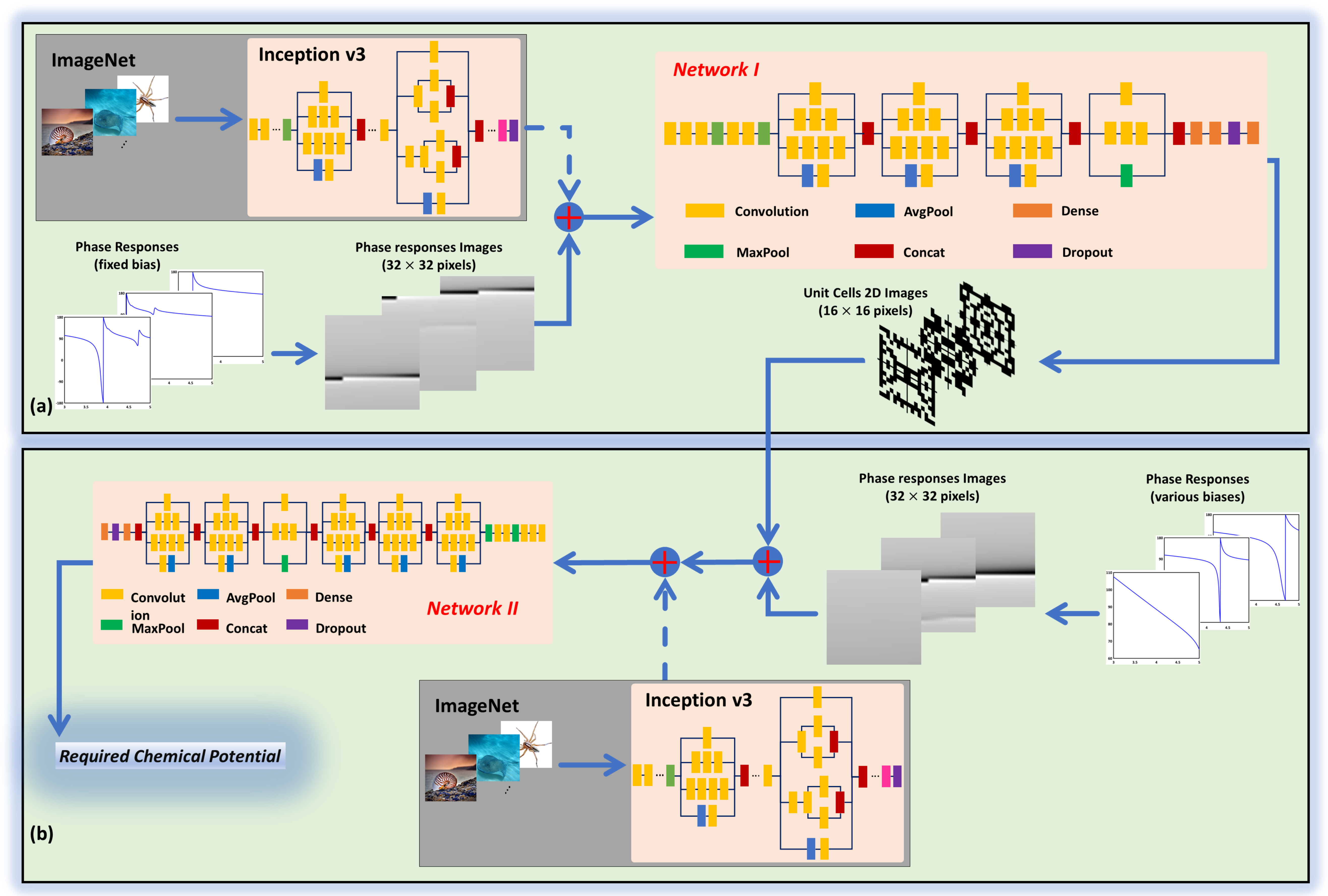}
    \caption{\label{fig:model_schematic} Schematic diagram of the proposed model for the inverse design of reconfigurable metasurfaces. (a) Network 1: inverse design of passive parts of graphene-based meta-atoms. (b) Network 2: prediction of required chemical potentials to achieve tunable behaviors from the passive meta-atoms.}
\end{figure*}
\section{Meta-atom structure}
    This study employs reflective meta-atoms, comprising five layers, to manipulate EM waves. Figure 1 shows the schematic diagram of these meta-atoms, which are designed to control EM waves effectively. The first layer of the meta-atoms contains randomly arranged vacuum and graphene square blocks with a length of $l=0.5$ $\mu m$. The second and third layers consist of ultra-thin alumina and silicon layers that serve the sole purpose of electrostatically biasing the graphene layers and have a negligible impact on the EM responses \cite{rouhi2019multi}. The other layers of the meta-atoms consist of a silicon dioxide layer ($\epsilon_{r}=2.1$) serving as the primary substrate of the metasurface, with a thickness of $h=2.0$ $\mu m$, and a very thin gold layer. The meta-atoms for practical applications are incorporated into lattices, to decrease corner-related coupling effects. They are linked together through horizontal and perpendicular graphene ribbons with width $w=0.1$ $\mu m$ to simplify the biasing of the meta-atoms in lattices and use one electrostatic bias for them. The reflective meta-atoms employed in this study possess a periodicity of $p=10$ $\mu m$ in both the x and y directions. The graphene-based layer is subdivided into 16$\times$16 square blocks and exhibits twofold symmetry along the x- and y-axis in the x-o-y plane. Consequently, the first layer of the meta-atoms can be efficiently represented using 8$\times$8 coding sequences.
	
	DL models rely on the initial data used to train the model. Even the most effective models may become worthless without a base of high-quality training data. Indeed, when trained on insufficient, incorrect, or irrelevant data, strong DL models can fail severely at generalization. Hence, to develop a high-performing model for the inverse design of reconfigurable graphene-based metasurfaces, a dataset composed of 8$\times$8 coding sequences, which represent the geometry of meta-atoms, graphene material electrostatic biases, and the corresponding phase responses are collected. The coding sequence of graphene material in the meta-atoms is randomly selected. Moreover, in order to preserve a balanced dataset, 10 chemical potentials between 100 $meV$ to 1000 $meV$ are chosen for each meta-atom structure. The first chemical potential is considered 100 $meV$, while the others are randomly chosen in 100 $meV$ intervals. For example, the last chemical potential is a random value between 900 $meV$ and 1000 $meV$. Figure 2 shows the distribution of the chemical potentials in different intervals. As clearly shown, the distribution has an almost uniform density in the whole 100 $meV$ to 1000 $meV$ period. The randomization of the meta-atoms in terms of graphene geometry and chemical potential significantly enhances the generalization performance of deep NNs for the inverse design of unseen EM responses in the training dataset. The EM responses of the generated meta-atoms are calculated by CST Microwave Studio (CST MWS) using the Frequency Domain Solver. In the CST MWS environment, periodic boundary conditions are activated in the x- and y-directions and Floquet ports are defined along the z-direction, resulting in the simulation of a transversely infinite array composed of graphene-based meta-atoms. Using the DL model, the reflection coefficients of metasurfaces are explored in order to design reconfigurable metasurfaces with desired EM functions, in the THz regime.
   \begin{figure*}[!t]
	\subfigure[]{%
	\label{fig:F4a}%
	\includegraphics[height=2.9 in]{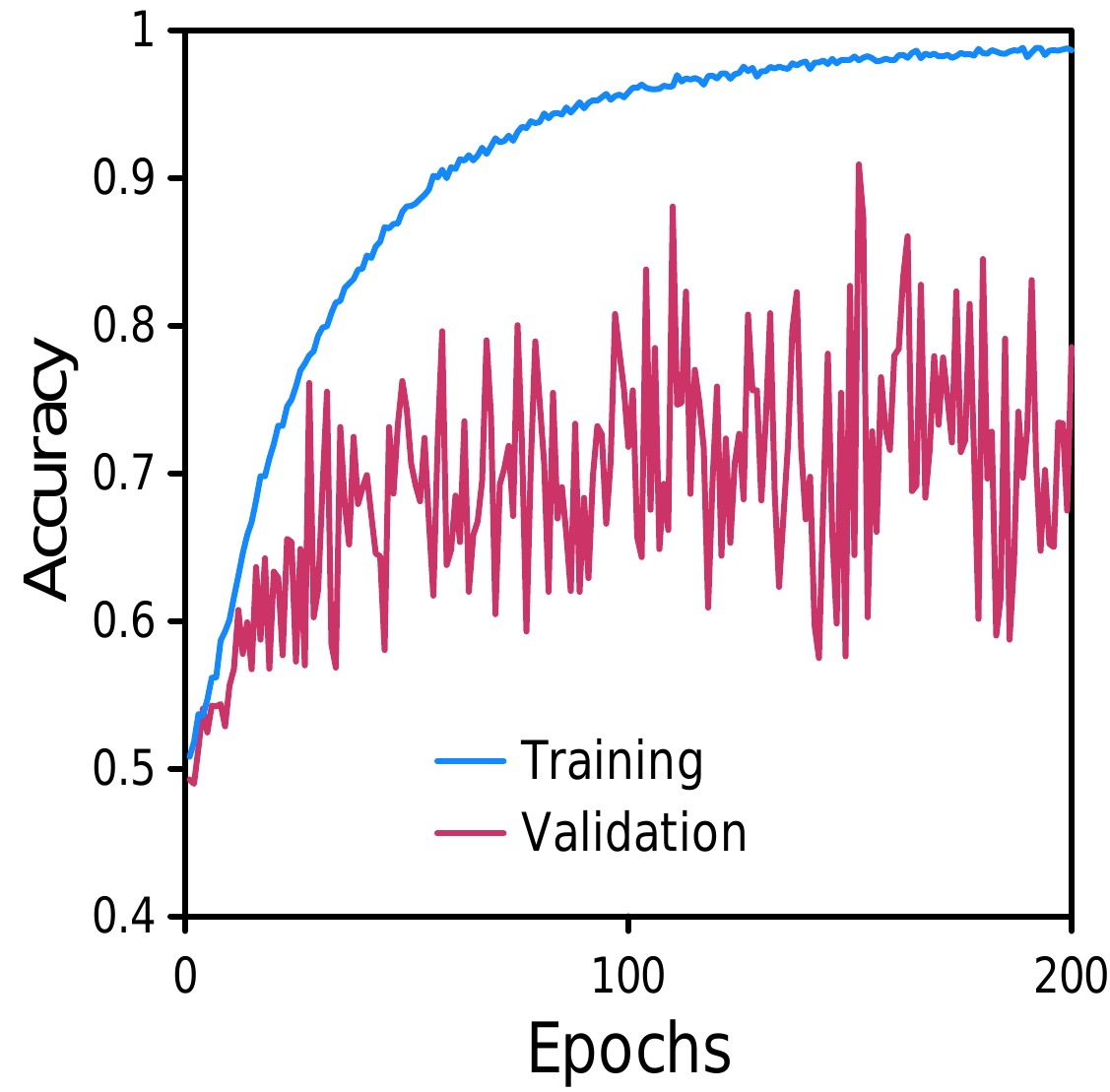}}%
	\subfigure[]{%
	\label{fig:F4b}%
	\includegraphics[height=2.9 in]{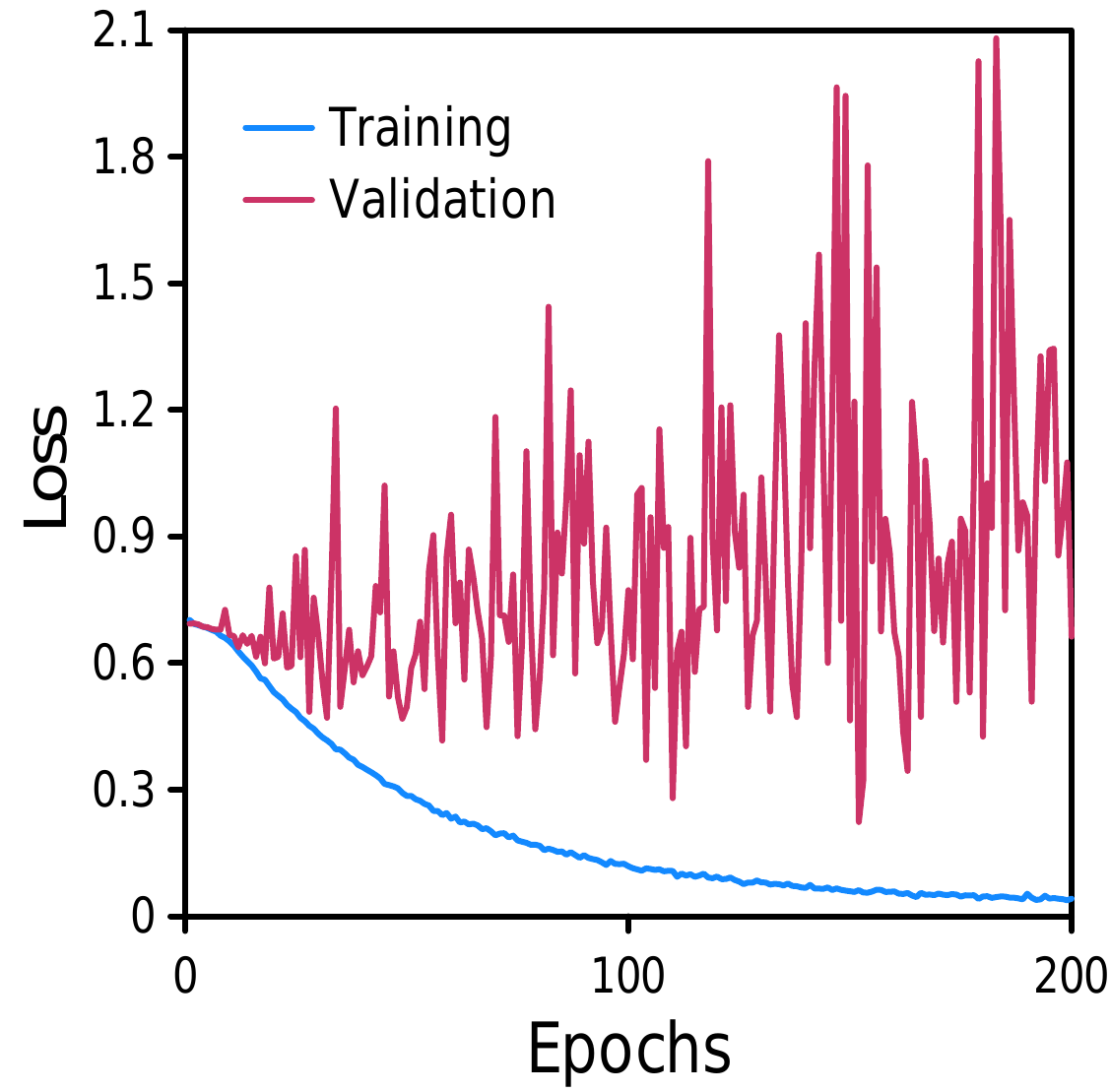}}%
        \qquad
	\subfigure[]{%
	\label{fig:F4c}%
	\includegraphics[height=2.9 in]{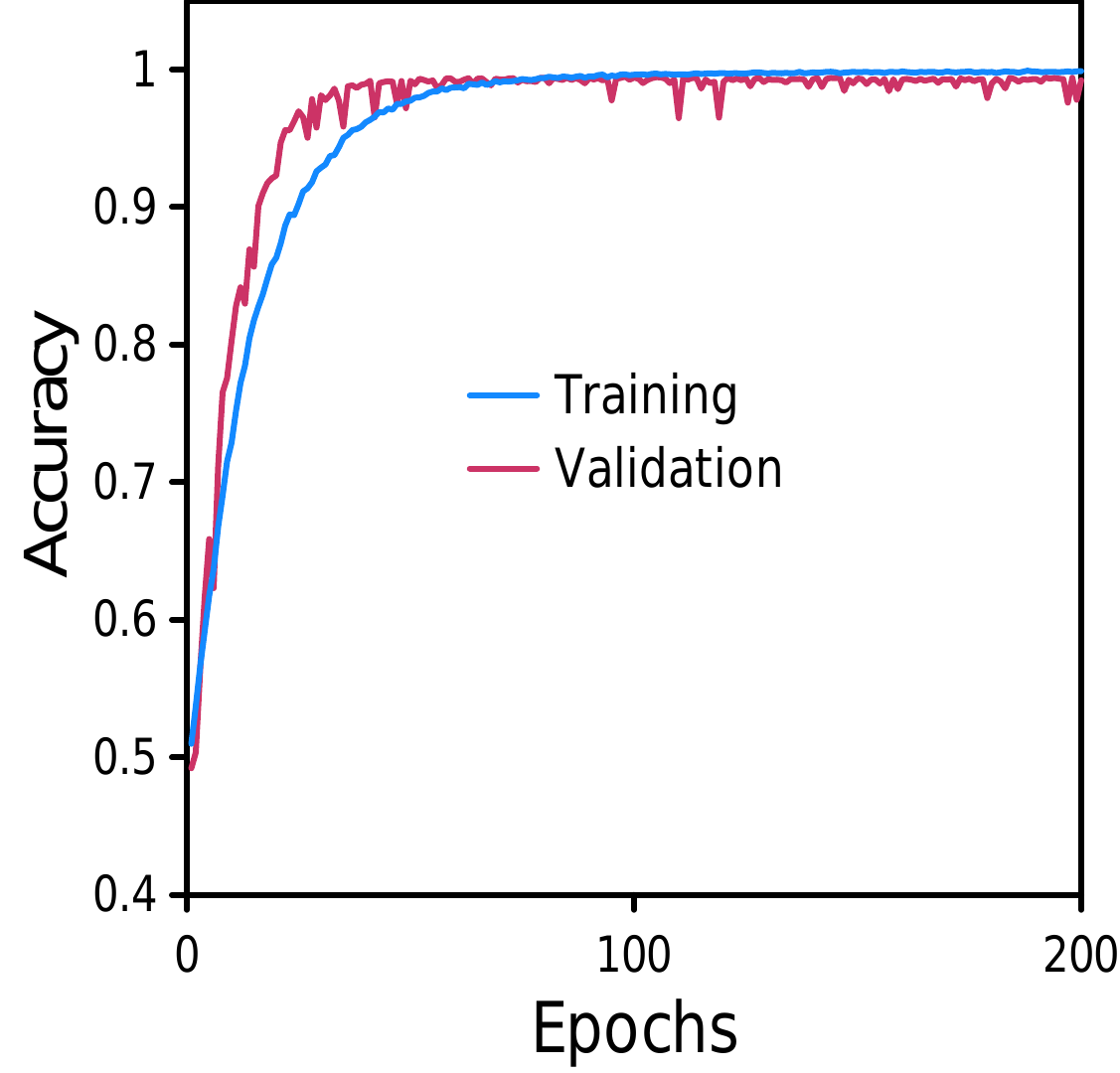}}%
	\subfigure[]{%
	\label{fig:F4d}%
	\includegraphics[height=2.95 in]{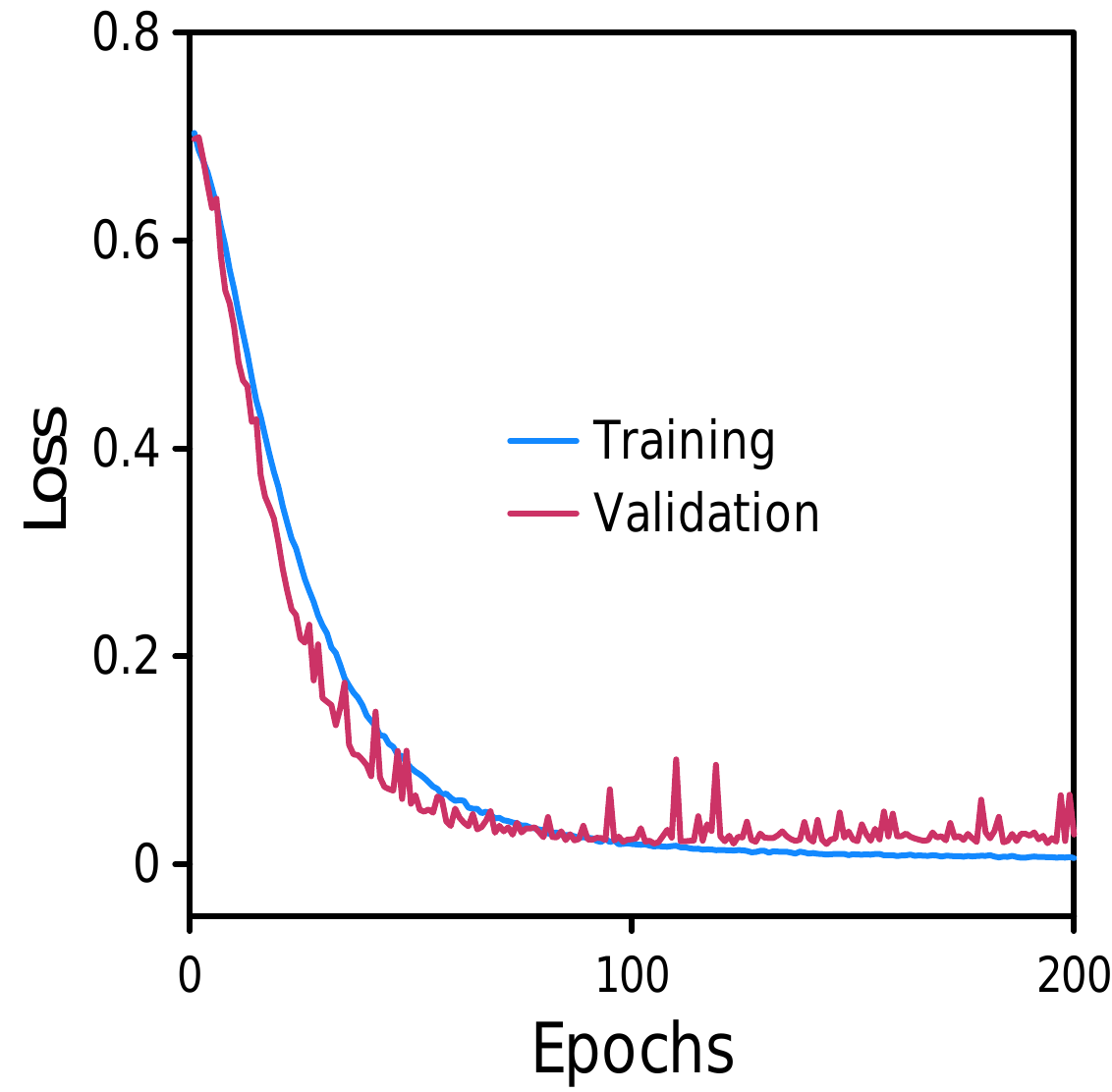}}%
	\caption{\label{fig:epsart} Performance comparison of Network 1, passive metasurface designer, during training and validation with and without TL. Training and validation (a) accuracy, (b) loss without TL, (c) accuracy, and (d) loss with TL.}
    \end{figure*}

    \section{Inverse design model}
    In 2016, Szegedy et al. suggested Inception v3, the third generation of Google Net, for assisting in image analysis and object detection \cite{szegedy2016rethinking}. The Inception v3 model is a highly robust and advanced CNN architecture that consists of 48 meticulously crafted layers. Its power lies in its innovative utilization of symmetric and asymmetric building blocks, including convolutions, average pooling, max pooling, concatenations, dropouts, and fully connected layers. To ensure stable training, batch normalization is extensively incorporated throughout the model and applied to activation inputs. With its intricate design, Inception v3 exhibits unparalleled accuracy and performance across various tasks, establishing itself as one of the most formidable models in the field of DL. Notably, it achieves an impressive 78.1\% accuracy on the extensive ImageNet dataset, which encompasses over 14 million images across more than 20,000 categories. 
    However, due to the processing of approximately 25 million parameters on a vast dataset, training Inception v3 demands substantial computational resources. This high cost makes it unsuitable for use in applications like metasurface design, where gathering a huge dataset is impossible. A way to short-cut this time-consuming process is to use TL. TL is a novel machine learning framework based on DL that provides for rapid progress or enhanced performance when modeling a new problem by transferring knowledge from a previously learned related task \cite{goodfellow2017deep}. TL methods attain optimal performance faster than standard DL models, which require training from scratch and a significant quantity of data. In this paper, a model based on TL is employed for the inverse design of reconfigurable graphene-based metasurfaces at the THz frequency regime. This model consists of two consecutive CNNs in which pre-trained Inception v3 models are used as the starting point for reconfigurable metasurfaces design. 

    Each tunable graphene meta-atom of the metasurface can be described with two matrices: the first matrix, 8$\times$8 matrix or a binary vector with 64 elements, represents the passive structure of the meta-atom; while the second one is a 1-element matrix that shows the chemical potential of the meta-atom. The reflection phase of the graphene meta-atom is studied in 1024 frequency points from 3 THz to 5 THz, so it can be represented by a 32$\times$32 matrix (image with 1024 pixels). The inputs of the inverse design model are the desired reflection phases and the output of the model is the tunable meta-atom (geometrical structure as well as chemical potentials). To achieve an advanced inverse design platform for reconfigurable graphene-based metasurfaces, a groundbreaking model composed of two CNNs is developed. The workflow of the inverse design platform is presented in Figure 3. 
    
    The first CNN is designed to take in 32$\times$32 images of reflection phases as input and produce 64-element binary vectors that represent the geometrical structure of the meta-atoms. Its purpose is to design the passive components of the meta-atoms while maintaining a fixed chemical potential of 100 $meV$. This choice of fixed chemical potential is made to ensure stable and controlled behavior for the passive meta-atoms, allowing the focus to be on manipulating the reflection phase by engineering the geometry. In the subsequent step, if a different reflection phase is desired for the meta-atom designed in the previous step, the second CNN comes into play. This network takes an image containing a desired reflection phase independent of the one inputted in Network 1, along with the passive meta-atom generated by the first network, and predicts the appropriate chemical potential needed to achieve the desired response. By learning the relationship between the reflection phase, geometrical structure, and the required chemical potential, the second CNN enables the inverse design of reconfigurable metasurfaces with the desired EM functions. It is important to highlight that the model's capability is not limited to designing meta-atoms that exhibit only two desired EM responses. Instead, it can be used to inverse design meta-atoms that demonstrate multiple desired EM responses by simply adjusting the chemical potential. The architectures and performances of both networks are analyzed in detail in the subsequent sections.
    \begin{table}
  \centering
  \caption{The performance of Network 1 for different numbers of training data samples.}
  \begin{tabular}{cc}
    \toprule
    \textbf{\# Samples} & \textbf{Correlation Distance ($ \% $)} \\
    \midrule
    6,000  & 24.8 \\
    12,000  & 95.4 \\
    18,000  & 96.2 \\
    27,000  & 97.8 \\
    \bottomrule
  \end{tabular}
\end{table}

\begin{table}
  \centering
  \caption{The performance of different DL-based models in developing Network 1.}
  \begin{tabular}{ccc}
    \toprule
    \textbf{Model} & \textbf{Trainable Parameters} & \textbf{Correlation Distance ($ \% $)}\\
    \midrule
    TL Inception V3  & 9,291,488 & 95.4\\
		      MobileNet & 7,467,904  & 54.2\\
		      DenseNet 121 & 8,069,056 & 59.5\\
		      DL Inception V3 & 9,291,488 & 52.4\\
    \bottomrule
  \end{tabular}
\end{table}

    \begin{figure*}[]
	\subfigure[][]{%
	\label{fig5a}%
	\includegraphics[height=2.85 in]{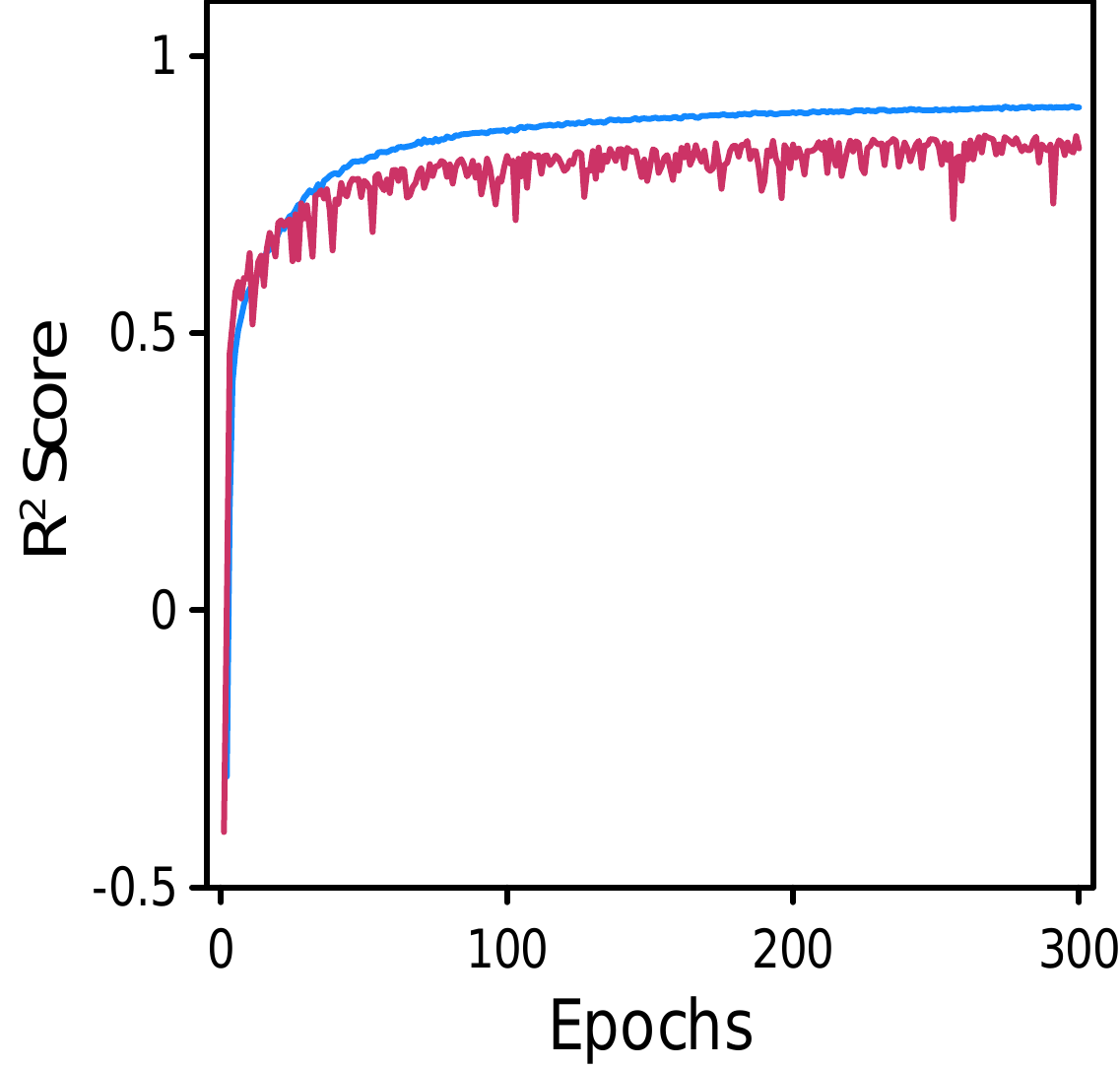}}%
	\subfigure[][]{%
	\label{fig5b}%
	\includegraphics[height=2.85 in]{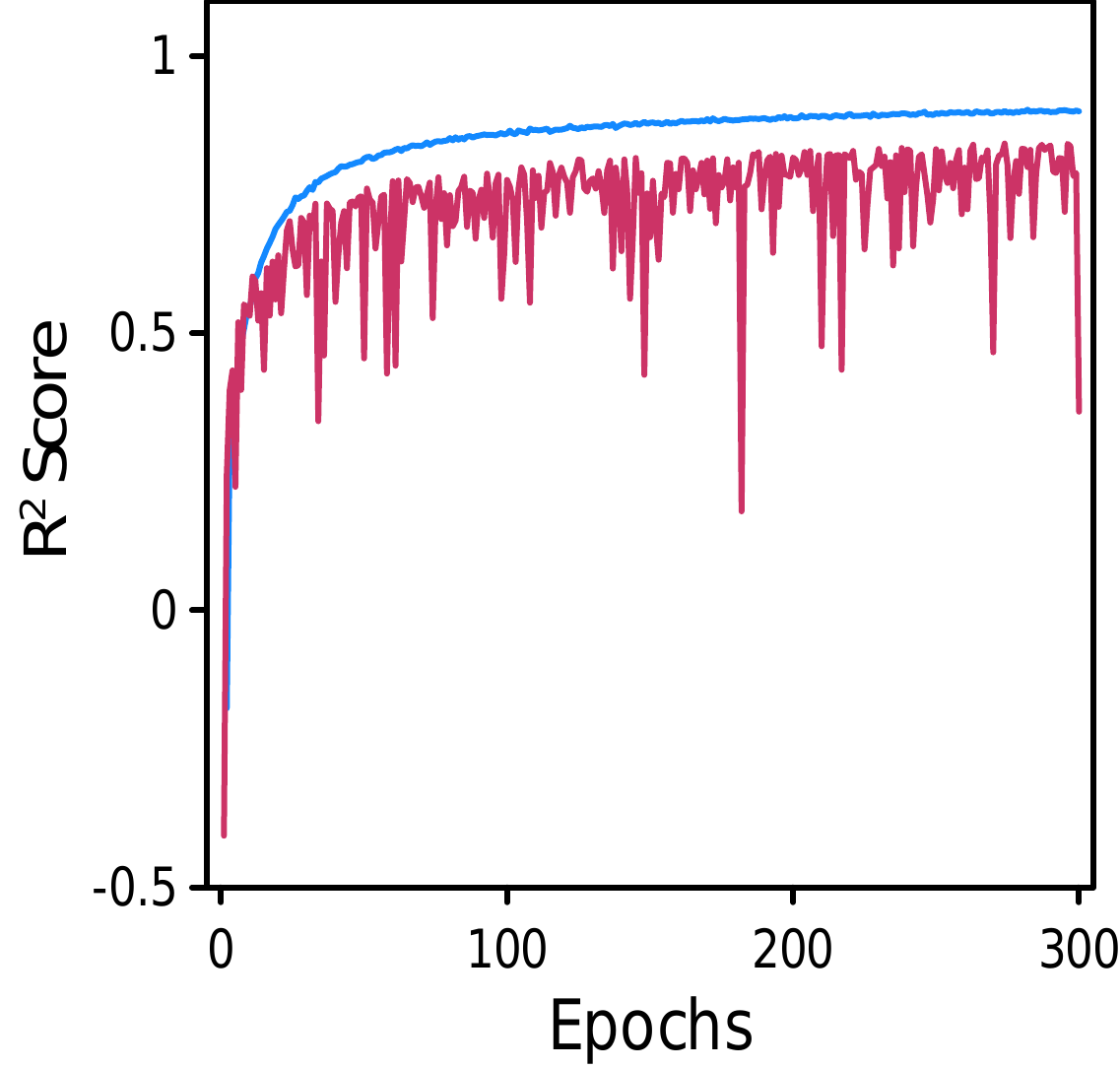}}%
	\caption{\label{fig:epsart} {Performance comparison of Network 2, chemical potential predictor, during training and validation with and without TL. Training and validation R2-score (a) without TL and (b) with TL. }}
    \end{figure*}
 
\subsection{Network 1, passive metasurface designer}
        In the proposed methodology, the design process of tunable graphene meta-atoms, which exhibit different  EM responses based on external control signals, is divided into two distinct parts. The first part involves designing the passive components of the meta-atoms, while the second part focuses on predicting the appropriate external control signals or chemical potential values. 

        To design the passive components of the tunable graphene meta-atoms, Network 1 is employed. The power of TL and CNNs is combined in this network to extract essential features from reflection phase images. The feature extractor in Network 1 utilizes the Inception v3 model up to the ``mixed 5" layer, which is pre-trained with weights from the ImageNet database. By utilizing this pre-trained model, the network benefits from the extensive knowledge and expertise gained through training on a large and diverse image dataset. The advanced Keras library is used to import the pre-trained Inception v3 model. This feature extraction step is crucial for capturing the important characteristics and patterns present in the reflection phase images, facilitating the design of the passive components of the meta-atoms. 

        The powerful Inception v3 network is initially utilized to extract essential information from images in the proposed methodology. However, to optimize the network for the specific task of estimating the passive structure of tunable meta-atoms, the uppermost layers of the Inception v3 network are replaced with a single trainable layer. This new layer is carefully designed with an impressive node count of 1024, allowing for efficient estimation of the passive structure.
        
        One potential challenge when training the new network is the risk of rapid overfitting, particularly when working with a limited number of training examples. To mitigate this issue and enhance the network's generalization ability, dropout regularization is implemented. Dropout regularization randomly deactivates a portion of the nodes during training, forcing the network to rely on the remaining nodes and preventing over-reliance on specific features. This regularization technique enhances the efficiency of the CNN in the inverse design of graphene meta-atoms that were not utilized during the model's training phase. The incorporation of this configuration, which combines Inception v3 with a trainable layer and dropout regularization, enables the training of the graphene meta-atom inverse design network with greater efficiency and effectiveness compared to traditional DL techniques, even when working with a limited amount of data.

        As shown in Figure 3(a), the initial stage of the design process involves inputting 32$\times$32 matrices or 1024-pixel images representing reflection phases into Network 1. This network generates corresponding 64-pixel graphene meta-atoms, where each element is either 0 or 1 to indicate vacuum and graphene blocks, respectively. Since the inverse design of the passive components of graphene meta-atoms is a multi-output classification task, Network 1 maps a single input image to 64 distinct binary outputs. To accomplish this, sigmoid functions are used to predict the probability of each binary variable in the multi-output binary classification. Accuracy and loss metrics are employed to evaluate the performance of Network 1 in designing these meta-atoms, measuring the classification model's accuracy and loss. However, pixel arrangements in the graphene meta-atoms are crucial, and there are strong correlations between different pixels within the meta-atoms. Therefore, these two metrics alone are insufficient to fully represent the network's performance.

        To address this issue, the correlation distance between the original vectors in the training and validation datasets and the generated output vectors is measured and considered as the evaluation metric for Network 1's performance. The correlation distance is a statistical measure of dependence between two vectors, defined as the distance between two vectors based on the absolute value of their pairwise correlations. The correlation distance between vectors u and v is defined as:

        \begin{equation}
           dCor = 1 - \frac{(u - \bar{u})(v - \bar{v})}{(|| u - \bar{u}||_{2}|| u - \bar{v}||_{2})}
        \end{equation}

        where $\bar{u}$ and $\bar{v}$ represent the mean of vectors $u$ and $v$, respectively. Incorporating this metric allows for a more accurate evaluation of Network 1's performance in designing passive graphene meta-atoms, ensuring a more robust and reliable design process.

       In order to optimize the trainable parameters in the TL-based CNN with respect to the correlation distance, the Adam optimizer is employed. The Adam optimizer is well-known for its rapid convergence speed and efficiency compared to traditional gradient descent methods. For this optimization, an initial learning rate of $1\times10^{-4}$ is utilized. Furthermore, the optimization process yields superior results when performed with a batch size of 32 over 200 epochs. These parameter choices have been carefully selected to ensure optimal performance and efficiency in optimizing the trainable parameters within the proposed TL-based CNN.

        \subsubsection{Network 1 evaluation}
            In this part, the performance of the first network of the proposed methodology in designing passive components of tunable meta-atoms is investigated. As previously stated, the pre-trained model is utilized as the starting point for a model on meta-atom inverse design to save training time, improve CNN performance, and avoid the requirement for a large amount of data. In this regard, the accuracy and loss of both the training and validation datasets are studied in two cases: 1- training the network when the starting points are random numbers, and 2- training the network when the starting points are the weights of the pre-trained Inception v3 using ImageNet. Figures 4(a) and 4(b) show the accuracy and loss of the case 1. Upon reviewing the learning curves during training, it becomes apparent that the model quickly becomes overfitted on the training data. Simultaneously, the validation loss demonstrates an increasing trend with significant spikes, while the validation accuracy remains stagnant with substantial fluctuations. As a result, in the case 1, although the model is very specialized in predicting the training dataset, it cannot generalize to new data samples. This problem is known as overfitting in data science. To alleviate the overfitting problem, a large amount of data samples should be added to the training dataset. However, for the metasurface design problem data collection is expensive, time-consuming, and difficult. In addition, the training time of the model with the larger training set will increase drastically. In the case 2, however, as seen in Figure 4(c), the training and validation accuracies increase to a point of stability with a minimal gap between the two final accuracy values. The training and validation losses likewise converge to a value from the loss standpoint (see Figure 4(d)).

            Table 1 provides a comprehensive evaluation of the performance of Network 1 based on correlation distance for different sizes of training datasets. The results in Table 1 show that when the training dataset size exceeds 12,000 samples, the correlation distance exceeds 95\%. Furthermore, the number of trainable parameters and correlation distance of several sample networks, including the Inception v3 TL network, MobileNet TL network, DenseNet 121 TL network, and Inception v3 DL network, are compared to demonstrate the superior performance of Network 1 (Inception v3 TL network) in training reflection phases. It is worth noting that MobileNet and DenseNet are alternative CNN architectures commonly used in various DL tasks, including image classification, object detection, and natural language processing \cite{huang2017densely, howard2017mobilenets}. As presented in Table 2, despite having the same architecture and trainable parameters, the Inception v3 DL and TL models exhibit vastly different performances in learning the mapping between the reflection phases and meta-atom structures. Specifically, the pre-trained weights of the Inception v3 TL network, which are used as starting points for Network 1, enable more efficient feature extraction, resulting in better dataset fitting. Conversely, the other two TL models, MobileNet TL network and DenseNet 121 like Inception v3 DL network fail to extract important features of the EM responses, indicating their performance inferiority in comparison to Network 1. These findings highlight the potential of the proposed methodology for designing and optimizing meta-atoms, as well as the importance of appropriate network architecture and training strategies.

 \begin{table}
  \centering
  \caption{The performance of Network 2 for different numbers of training data samples.}
  \begin{tabular}{ccc}
    \toprule
    \textbf{\# Samples } & \textbf{Training $R^2$ score} & \textbf{Validation $R^2$ score}\\
    \midrule
    10,000  & 0.898  & 0.803\\
                  20,000  & 0.900 & 0.808\\
                  30,000  & 0.899 & 0.833\\
                  36,000  & 0.905 & 0.870\\
    \bottomrule
  \end{tabular}
\end{table}

 \begin{table}
  \centering
  \caption{The performances of different DL-based models in developing Network 2.}
  \begin{tabular}{cccc}
    \toprule
    \textbf{Model} & \textbf{Trainable Parameters} & \textbf{Training $R^2$ score} & \textbf{Validation $R^2$ score}\\
    \midrule
    TL Inception V3  & 6,977,825 & 0.905 & 0.870 \\
		         MobileNet  & 5,305,153 & 0.900 & 0.652 \\
		         DenseNet 121  & 7,479,169 & 0.907 & 0.854 \\
		         DL Inception V3  & 6,977,825 & 0.900 & 0.358 \\
    \bottomrule
  \end{tabular}
\end{table}

\begin{figure}[]
	\centering
	\includegraphics[height=3.3in]{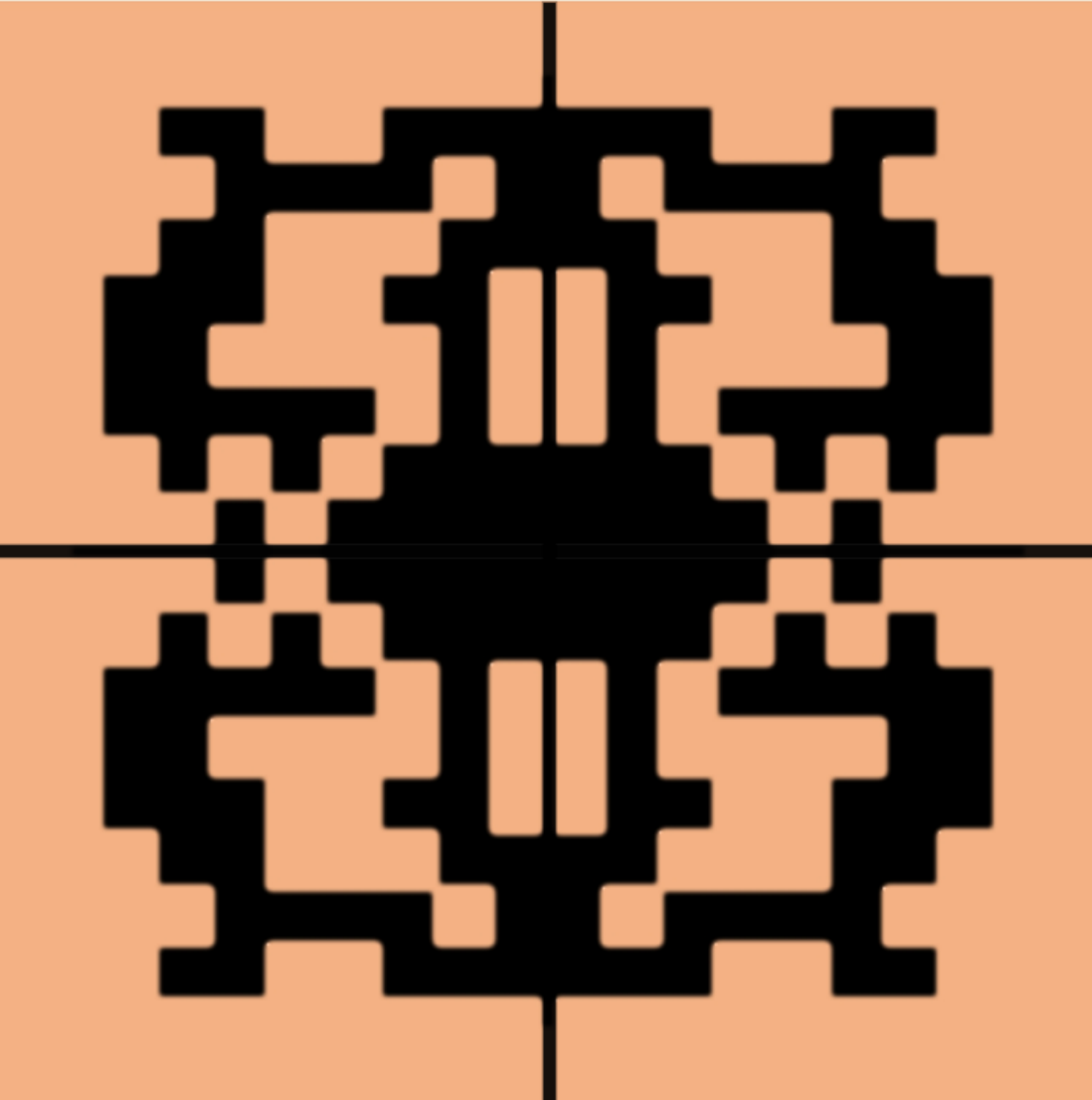}
	\caption{\label{fig:epsart} {The passive structure of tunable meta-atom designed by the TL-based inverse       design model.}}
\end{figure}

   \subsection{Network 2, chemical potential predictor}
        The tunable properties of graphene meta-atoms make them highly desirable for use in reconfigurable metasurfaces. However, to achieve the desired EM responses from these meta-atoms, the chemical potentials must be carefully designed. This is where Network 2, based on TL and employing a mapping between the reflection phase and chemical potentials, comes in. As can be seen in Figure 3(b) using a set of 1088-pixel images composed of the meta-atoms designed by Network 1 and the desired reflection phases, Network 2 predicts the chemical potentials required for optimal tunability. By doing so, it identifies the ideal chemical potential match for obtaining the tunable response desired from the designed passive meta-atoms of Network 1. To accomplish this task, Network 2 utilizes Inception v3 up to layer “mixed 7” pre-trained with ImageNet as the feature extractor, with one fully-connected trainable layer containing 256 nodes. Dropout regularization is also employed to prevent overfitting. The training process of Network 2 involves the Adam optimizer with an initial learning rate of $1\times10^{-4}$, which adapts the learning rate dynamically during training and incorporates momentum and adaptive learning rate scaling.

   \begin{figure*}[!t]
 
	\subfigure[]{%
	\label{fig:F7a}%
	\includegraphics[height=2.9 in]{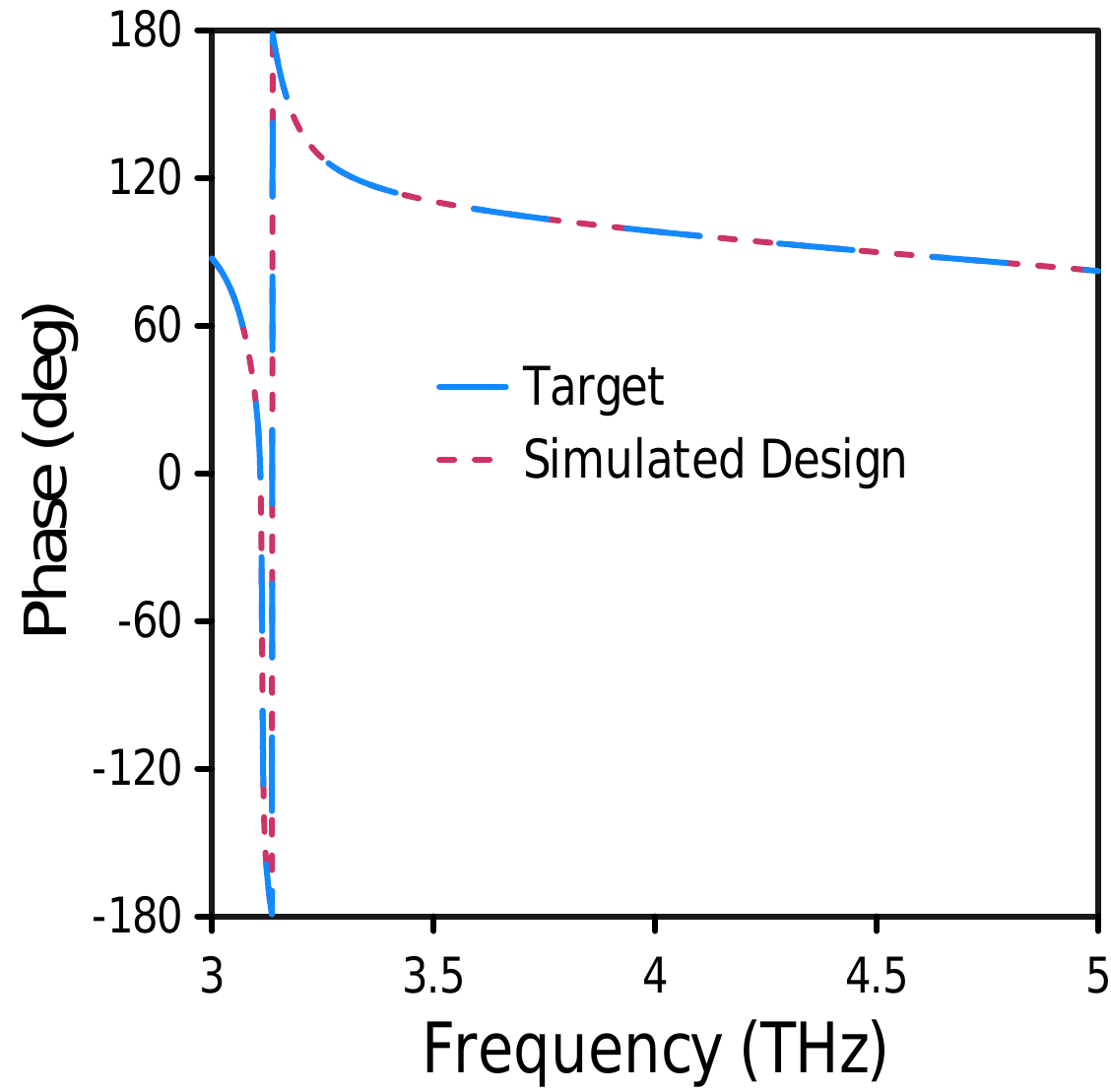}}%
	\subfigure[]{%
	\label{fig:F7b}%
	\includegraphics[height=2.9 in]{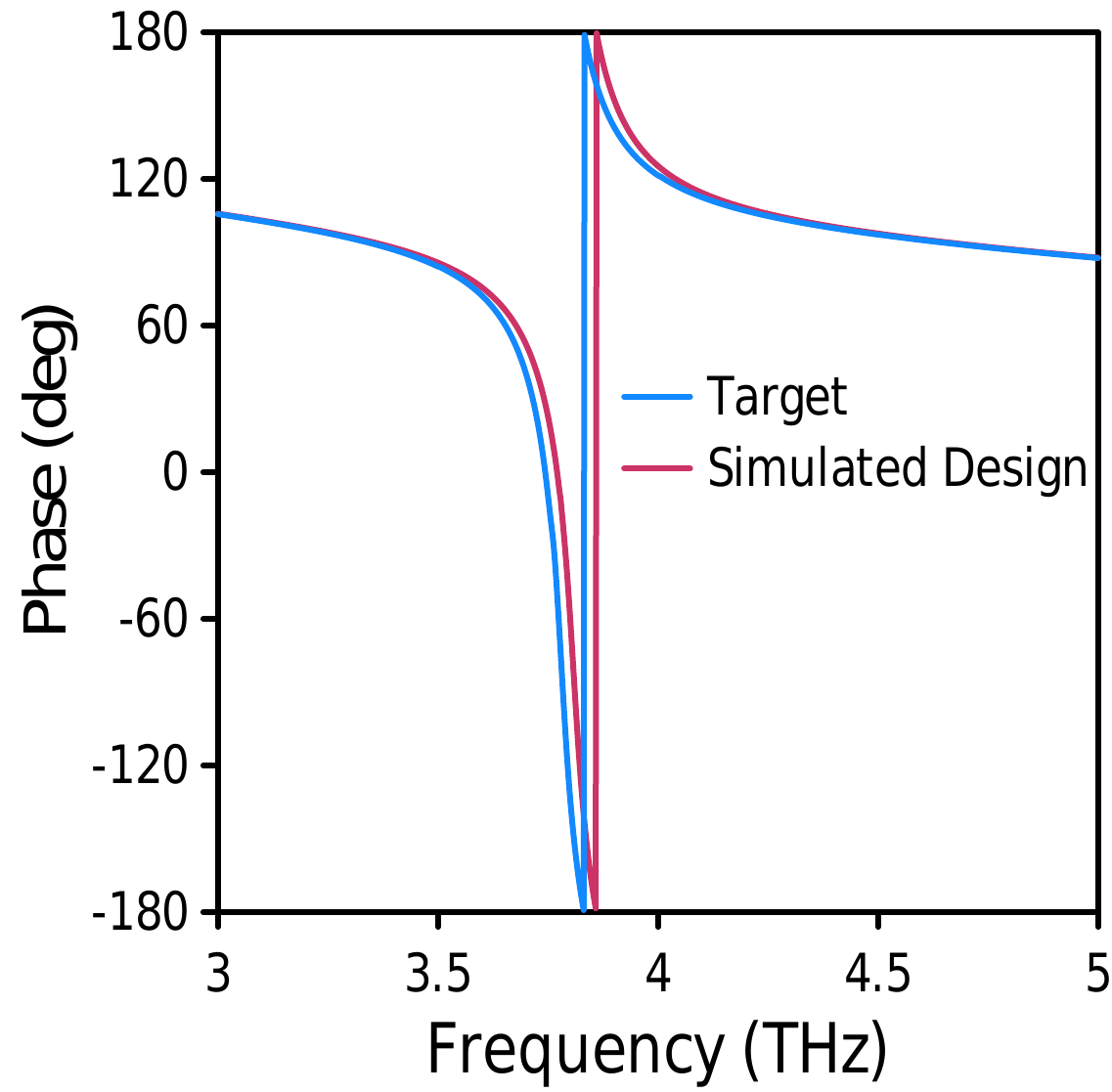}}%
        \qquad
	\subfigure[]{%
	\label{fig:F7c}%
	\includegraphics[height=2.9 in]{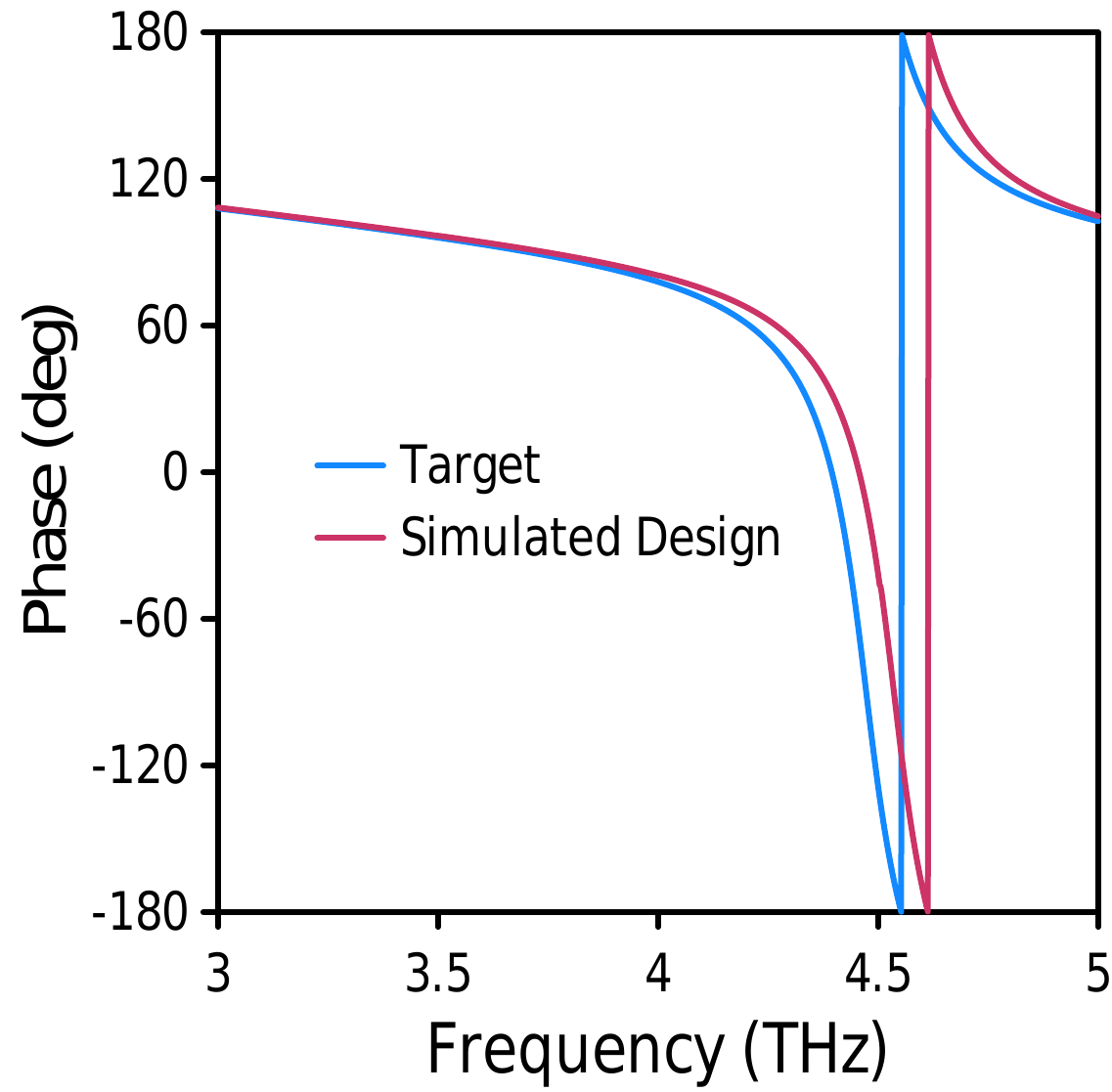}}%
	\subfigure[]{%
	\label{fig:F7d}%
	\includegraphics[height=2.9 in]{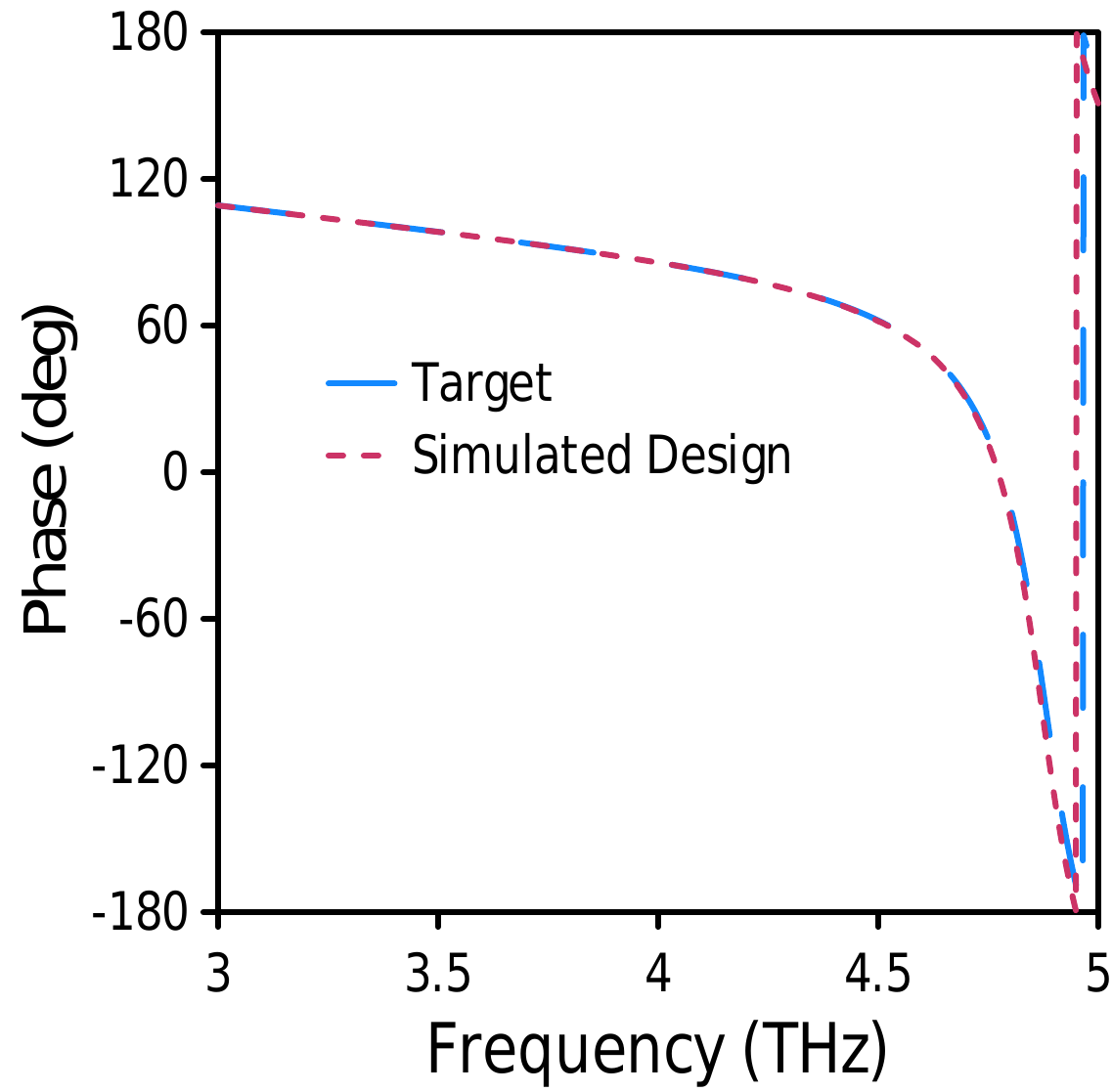}}%
	\caption{\label{fig:epsart} The target and simulated reflection phases of the tunable meta-atom designed 
         by the inverse design model at the designed chemical potentials. (a) 100 $meV$ (b) 148 $meV$ (c) 225 $meV$ (d) 270 $meV$.}
    \end{figure*}

        \subsubsection{Network 2 evaluation}
            Building and deploying a successful regression model requires the selection of an appropriate metric for evaluating its performance, which can enable better performance optimization and fine-tuning and ultimately leads to improved results. While there are several metrics available to evaluate regression models, such as Mean Square Error (MSE), Root Mean Squared Logarithmic Error (RMSLE), and $R^2$ score, the $R^2$ score metric stands out as an effective metric for evaluating regression models' performance. This is because the $R^2$ score accurately reflects the proportion of variation in real values captured by the model, providing a more precise and meaningful assessment of its quality on a scale of $<$1. Adopting the $R^2$ score as the primary evaluation metric for regression models can significantly enhance the model's effectiveness, thereby leading to greater real-world impact and improved outcomes. The following relationship is used to express the $R^2$ score: 
            \begin{equation}
               R^2(y, \hat{y}) = 1 - \frac{\sum_{i=1}^{n}(y_i - \hat{y})^{2}}{\sum_{i=1}^{n}(y_i - \bar{y})^{2}}
            \end{equation}
            where $y_i$ is the actual output value of the $i$th sample in the dataset, $\hat{y}_i$ is the predicted output value of the $i$th sample by the regression model, $\bar{y}$ is the mean of the output values in the dataset, and $n$ is the total number of samples in the dataset.
        
            Figure 5 shows the $R^2$ score of Network 2 during training, both with and without TL. In Figure 5(a), the $R^2$ score of Network 2 based on Inception v3 with pre-trained weights from ImageNet is displayed. The training and validation $R^2$ scores converge to approximately 0.89 and 0.8, respectively, with a generalization gap. Despite this gap, these scores are acceptable for the task, especially considering that the network is trained with only 10,000 samples. Figure 5(b) explores the training process of Network 2 without using TL. The training $R^2$ score achieves similar stability as the TL-based network, but the validation $R^2$ score exhibits large spikes, indicating the limited size of the training dataset. However, increasing the size of the training dataset is challenging and requires time-consuming computations for the reconfigurable metasurface design problem. Hence, utilizing TL to predict the necessary chemical potentials of tunable graphene meta-atoms significantly reduces computing costs. 
            
            Additionally, the comparison of the TL network's performance in predicting the chemical potential of tunable graphene-based meta-atoms, considering varying numbers of training data samples, is presented in Table 3. The results reveal that as the number of training samples increases, the model's ability to generalize to unseen data from the same distribution improves. Remarkably, even with a relatively small dataset of 10,000 samples, the network achieves a satisfactory $R^2$ score of approximately 80\%. Interestingly, when the dataset is expanded threefold, the score only shows a marginal increase of 7\%. This highlights the diminishing returns associated with further increasing the dataset size.
    
            The effectiveness of Network 2 in predicting chemical potentials is further demonstrated by comparing the number of trainable parameters and $R^2$ scores of various sample networks, including the Inception v3 TL network, MobileNet TL network, DenseNet 121 TL network, and Inception v3 DL network (Table 4). Interestingly, despite having the same architecture and number of trainable parameters, the DL and TL models of Inception v3 exhibit vastly different performances in predicting the chemical potential of the designed meta-atoms for the desired EM responses. This suggests that integrating pre-trained layers to extract features from EM response images significantly enhances the architecture's efficiency in designing tunable meta-atoms. Moreover, while the pre-trained Inception v3 network is capable of deducing the salient features from the input images and producing satisfactory results, the other pre-trained models, MobileNet and DenseNet 121 networks are unable to achieve comparable performance. Therefore, the effectiveness of pre-trained models in designing tunable meta-atoms depends heavily on the specific architecture of the models.

\section{Discussion}
    A reconfigurable graphene metasurface is presented in this section, utilizing the proposed TL-based model to test its efficiency and performance in inverse design. Four different reflection phases are considered as the target EM responses, and the corresponding tunable meta-atom is designed using the inverse design model. Figure 6 shows the pixelated structure of the graphene layer. This designed graphene meta-atom exhibits different phase resonances at different frequencies, depending on the applied chemical potential. The methodology developed for achieving the corresponding tunable meta-atom to the desired reflection phases consists of two steps. In the first step, Network 1 designs the passive part of the graphene meta-atom to exhibit the first desired reflection phase at a chemical potential of 100 $meV$. In the second step, Network 2 predicts the required biases (chemical potentials) to reach the other three different phase responses using the meta-atom designed by Network 1. The predicted chemical potentials to realize the desired reflection phases are shown in Table 5. The designed tunable meta-atom is then re-entered into CST MWS, and its reflection phases at the predicted chemical potentials are studied and compared with the target ones. It can be seen in Figure 7(a) that the simulated phase response of the tunable meta-atom designed by Network 1 is in total agreement with the target phase response, and the other simulated reflection phases are also in good agreement with the target ones, demonstrating the excellent performance of the methodology in the inverse design of reconfigurable metasurfaces (Figure 7 (b)-(d)).

    \begin{table}
  \centering
  \caption{The predicted chemical potentials of the graphene meta-atom of Figure 6 to realize the target reflection phases.}
  \begin{tabular}{cc}
    \toprule
    \textbf{Sample} & \textbf{Chemical Potential ($meV$)}\\
    \midrule
    \#1  & 100 \\
		\#2  &148 \\
		\#3  & 225 \\
		\#4  & 270 \\
    \bottomrule
  \end{tabular}
\end{table}
\section{Conclusion}
    In this paper, a DL-based approach is presented to solve the inverse problem of designing reconfigurable graphene metasurfaces based on the desired reflection responses. The proposed methodology leverages two distinct CNNs to the inverse design of graphene metasurfaces that present several EM functionalities with real-time reconfigurability. The developed CNNs are trained using TL which significantly mitigates the computational costs of gathering training datasets. The pre-trained Inception v3 by ImageNet database is applied in the CNNs as the feature extractor to transfer the knowledge of image recognition to tunable metasurface design. The first network of the methodology designs the passive components of graphene meta-atoms, while the second network predicts the required chemical potentials of the reconfigurable graphene meta-atoms.
    
    The fully trained CNNs demonstrate their efficiency and capability in designing reconfigurable graphene meta-atoms via the design of a tunable metasurface at the terahertz regime. The results of numerical EM case study simulations confirm that the proposed TL-based methodology can be adopted as an efficient tool for designing tunable metasurfaces. This design technique is expected to be easily extended beyond graphene meta-atoms to other types of reconfigurable intelligent surfaces operating at multiple frequency bands suitable for next-generation wireless networks.
    
\printbibliography
\end{document}